\documentclass[preprint,aps,amsmath,amssymb,showpacs,superscriptaddress]{revtex4-1}
\usepackage{graphicx}
\usepackage{enumerate}
\usepackage{epsfig,amsmath}
\usepackage{subfigure}
\usepackage{hyperref}
\usepackage{relsize}
\usepackage{lmodern}
\usepackage{slantsc}
\usepackage{color}
\usepackage[usenames,dvipsnames]{xcolor}

\newcommand{\Ne}{\left<N_e\right>}
\newcommand{\IB}{\left<I_B\right>}
\newcommand{\IE}{\left<I_E\right>}
\newcommand{\RNum}[1]{\uppercase\expandafter{\romannumeral #1\relax}}

\begin{document}

\title{Dynamics of Bloch oscillating transistor near bifurcation threshold}

\author{Jayanta Sarkar}
\affiliation{Low Temperature Laboratory,  O.V.Lounasamaa Laboratory,\\ Aalto University, P.O. Box 13500, FI-00076 AALTO, Finland}
\author{Antti Puska}
\affiliation{Low Temperature Laboratory,  O.V.Lounasamaa Laboratory,\\ Aalto University, P.O. Box 13500, FI-00076 AALTO, Finland}
\author{Juha Hassel}
\affiliation{VTT Technical Research Centre of Finland, P.O. Box 1000, FI-02044 VTT, Finland}
\author{Pertti J. Hakonen}
\affiliation{Low Temperature Laboratory,  O.V.Lounasamaa Laboratory,\\ Aalto University, P.O. Box 13500, FI-00076 AALTO, Finland}

\date{\today} 

\begin{abstract}

Tendency to bifurcate can often be utilized to improve performance characteristics of amplifiers or even to build detectors. Bloch oscillating transistor is such a device. Here we show that bistable behaviour can be approached by tuning the base current and that the critical value depends on the Josephson coupling energy $E_J$ of the device. We demonstrate record-large current gains for device operation  near the bifurcation point at small $E_J$. From our results for the current gains at various $E_J$, we determine the bifurcation threshold on the $E_J$ - base current plane. The bifurcation threshold curve can be understood using the interplay of inter- and intra-band tunneling events.

\end{abstract}

\maketitle

In small Josephson junctions, charge and phase reveal
their conjugate nature in several macroscopic phenomena. The quantum nature of the
phase variable ($\varphi$) was shown in macroscopic tunneling experiments
\cite{clarkeprl}, while its conjugate relationship to the charge has
been shown in many consequent studies \cite{Tinkham1996}.
One of the consequences of the charge-phase conjugate relationship
is the Coulomb blockade of Cooper pairs which arises in ultra small
Josephson junctions having a capacitance ({\it C}) in the femtoFarad range \cite{haviland1,kuzmin1}. 
Charging energy and the Josephson coupling energy $E_J(\varphi)=-E_J\cos\varphi$ are the competing energy scales associated with these two variables.
 Accordingly, the Hamiltonian for the
small Josephson junction contains a periodic potential, and hence, Bloch
states with band structure appear. These bands are analogous to the conduction electron energy states in solid state
physics \cite{AL,LZ}.

The Bloch Oscillating transistor is a three-terminal mesoscopic device which is
based on the dynamics of the Bloch bands in a voltage biased
Josephson junction (JJ) in a resistive environment \cite{delahaye2,hassIEEE}. 
 The operation is due to an
interplay of coherent Josephson phenomena and Coulombic blockade of charge
transport which is controlled by single electron tunneling events.
The device can be viewed as a
charge converter of single electrons, induced from the base electrode, in to a sequence of $N$ sequential Cooper
pair tunneling events, \emph{i.e.,} Bloch oscillations on the emitter terminal
with a Josephson junction. The current gain
is ideally given by $\beta = 2N+1$. The number of Bloch oscillations
is limited by inter-band transitions caused by Landau-Zener (LZ) tunneling
which depends exponentially on the band gap between the ground and excited states of the Josephson junction.
This simple picture has been found to
correspond quite well to the measured current gain \cite{hassel04a}.

Incoherent tunneling of Cooper pairs and electrons, however, complicates the basic
BOT operation. The interaction of tunneling electrons or
Cooper pairs with the electromagnetic environment, has been demonstrated to be strong
in small tunnel junctions, both in the normal and
superconducting states \cite{devoret90,holst94}. Inelastic effects may,
for example, limit the lifetime of the Coulomb blockaded state
and, consequently, bias-induced changes in the inelastic tunneling rates can cause large
modifications in the operating point, and thereby contribute to the current gain of the BOT. These effects, in fact, are the foundation for bifurcation in the BOT operation because they allow the existence of two steady states at a fixed base current $I_B$. The existence of a bifurcation point is important as, with proper design, the vicinity of such a point can be employed to improve the characteristics of the BOT. In this paper, we investigate experimentally the bifurcation threshold in the BOT, and demonstrate record large current gains for small-$E_J$-device operation near the threshold. From our results for the current gains at various $E_J$, we determine the bifurcation threshold curve on the $I_B$ -- $E_ J$ plane. The measured transition curve can be qualitatively explained using a simple analytic approach, in which intra-band transitions are taken in to account phenomenologically, together with the transition rates due to inelastic tunneling.

This paper is organized as follows. In Sec. I, we first outline the basic
principles for understanding the electron tunneling dynamics in a
Bloch oscillating transistor. We will concentrate to the dynamics near the
bifurcation point at which the current gain of the device diverges. Our analytic model is verified using numerics with a similar approach as done in Refs. \cite{hassel04a,hassIEEE,hassel04}.
 Sec. II will describe sample manufacture
and experimental measurement techniques. Experimental results are presented in Sec. III. We will present data on the current gain at various values of Josephson energy, and construct a curve for bifurcation threshold on $E_J$ vs base current plane. The relation of current gain with the distance from the bifurcation point is also studied in detail. In Sec. IV, we discuss our results in the light of analytical and numerical calculations. 

\section{Theory}
\subsection{Band model of mesoscopic Josephson junctions}

In mesoscopic tunnel junctions, the discreteness of charge starts
to play a role via the Coulomb energy
$E_{\raisebox{-2pt}{\mdseries\itshape \scriptsize C}}=\frac{Q^{2}}{2C}$, where $C$ is the capacitance of
the junction and $Q$ is the charge on the capacitor plates. In
quantum theory, charge is described by the operator
$\widehat{Q}=-i2e\frac{\partial }{\partial \varphi }$, where $\varphi$ denotes the phase difference of the order parameter fields across the junction. This
operator is canonically conjugate to $\hat{\varphi} $, {\it i.e.,}
$[\hat{Q},\hat{\varphi}]=i2e$. Hence, there is a Heisenberg
uncertainty relation, $\Delta Q\Delta \varphi \sim 2e$, which
implies that the charge and the phase of the superconducting
junction cannot be defined simultaneously. This leads to 
delocalization of the phase and to Coulomb blockade of the
supercurrent, as experimentally shown by Haviland {\it et al.}
\cite{haviland1} in the case when Josephson energy is on the order
of the single-electron Coulomb energy, {\it i.e.,} $E_J/E_C$
$\sim 1$. The same conclusion of
delocalization applies even for large values of the ratio
$E_J/E_C$ \cite{Schon}.

Using the differential operator due to the
commutation relation, we can immediately write
the quantum mechanical Hamiltonian \cite{LZ} as
\begin{equation}
H = -E_C \frac{\partial^2}{\partial (\varphi/2)^2} - E_J \cos{\varphi} .
    \label{Hamiltonian}
\end{equation}
When $E_C \gg E_J$,  charge is
a good quantum number, which leads to Coulomb blockade of Cooper
pairs and a complete delocalization of the phase. Equation
\ref{Hamiltonian} then takes the form of the Mathieu
equation with the well-known solutions of the form
$\Psi_n^q(\varphi) = e^{i\varphi q/2e}u_n(\varphi)$, where
$u_n(\varphi)$ is a $2\pi$-periodic function and the wave
functions are indexed according to band number $n$ and quasicharge
$q$. 
Verification of the existence of the energy bands has been carried
out by different methods \cite{prance92,flees97,lindell03}.

Voltage over the junction is given by $V=\frac{\partial E}{\partial q}$
which changes along the energy band when quasicharge is varied. Thus, to have current flowing in the junction, the bias voltage $V_C$ (on the collector, cf. Fig. \ref{BOTschema}) has to be larger than the maximum Coulomb blockade voltage of the lowest band $E_0$: $V_C>\frac{\partial E_0}{\partial q}|_{\textrm{max}}$.
If the  current through the junction is low enough, ${\rm{d}}q /
{\rm{d}}t \ll e\delta E_1/\hslash$, where $\delta E_1$ is the gap
between the first and second band, the quasicharge $q$ is
increased adiabatically and the system stays in the ground band.
The junction is then in the regime of Bloch oscillations; the voltage over the
junctions oscillates and Cooper pairs are tunneling at the
borders of the Brillouin zone, \emph{i.e.,} here at $q = \pm e$. Consequently, the current
through the junction is coherent and the voltage and charge over
the junction oscillate with the Bloch oscillation frequency
\begin{equation}
    f_B = I/2e,
    \label{Blochfreq}
\end{equation}
where $I$ is the current through the Josephson junction.

If the current $I$ is not adiabatically small,
we can have Zener tunneling between adjacent energy bands. The
tunneling is vertical, \emph{i.e.,}  the quasicharge does not change. The
probability of Zener tunneling between bands $n-1$ and $n$ when
$E_C \gg E_J$ is given by
\begin{equation}
    P^Z_{n,n-1} = \exp \left (-\frac{\pi}{8} \frac{\delta E_n^2}{n
    E_C} \frac{{\rm{e}}}{\hslash I} \right ) = \exp \left (
    -\frac{I_Z}{I} \right ),
    \label{zener}
\end{equation}
where $\delta E_n = E_n - E_{n-1}$ and $I_Z$ is the Zener breakdown current \cite{zener145and137,ben-jacob1,ben-jacob2,ben-jacob3}. Provided that $V_C<\frac{\partial E_1}{\partial q}|_{\textrm{max}}$ for the excited state $E_1$, the junction will become Coulomb blockaded on the band $E_1$ after a Zener tunneling event,
and no current will flow through it any more. The role of the third
terminal is to relax the Josephson junction back to the ground state
where a new sequence of Bloch oscillations can be started.

\subsection{Incoherent tunneling processes}

The external environment gives rise to current
fluctuations that couple linearly to the phase variable. These can
cause both up- and downwards transitions. The amplitude of the
fluctuations is given by the size of the impedance: the larger the
impedance the smaller are the current fluctuations and the
transition rates. As we will see later on, the successful
operation of the BOT requires one to control both the upwards and
downwards transition rates.
 When modeling the BOT analytically, we will make use
of the Zener transition rates and transitions due to charge
fluctuations, both derived in Ref.~\cite{Schon}.

The electromagnetic environment around tunnel
junctions affects the tunneling process by allowing exchange of
energy between the two systems \cite{devoret90,averin,girvin,IN}. The
influence of the external circuit can be taken into account perturbatively, for
example, using the so called $P(E)$-theory \cite{IN}.
A perturbative treatment of the Josephson coupling term gives rise
to a result for incoherent Cooper pair tunneling
\cite{averin,IN} where the tunneling electron rate is directly
proportional to the probability of energy exchange with the
external environment governed by the $P(E)$ function. Taking both positive and negative energy exchange into account, tunneling both inward and outward direction leads to the total current
\begin{equation}
        I(V) =\frac{\pi {e} E_J^2}{\hslash} \left (P(2eV)  - P(-2eV)
    \right ).
        \label{CPcurrent}
\end{equation}
The function $P(E)$ can be written as
\begin{equation}
    P(E) = \frac{1}{2 \pi \hslash} \int_{-\infty}^{\infty}
    {\rm{d}}t \exp \left [J(t) + \frac{i}{\hslash} E t \right ],
    \label{PE}
\end{equation}
which is the Fourier transform of the exponential of the phase-phase
correlation function
\begin{equation}
    J(t) = \left < \left [\varphi(t)-\varphi(0) \right ] \varphi(0)
    \right>.
    \label{phasephase}
\end{equation}
The phase-phase correlation function is determined by the
fluctuations caused by the environment and it can be related to the
environmental impedance via the fluctuation-dissipation theorem.

For a high-resistance environment, the $P(E)$ function is strongly peaked
at energies around $E_C$, and it may be
approximated by a Gaussian function
\begin{equation} \label{pe}
 P(E)=\frac{1}{\sqrt{4 \pi E_C k_B T}}\exp \left[-\frac{(E-E_C)^2}{4E_C k_B T}\right],
\end{equation}
where the width is governed by thermal fluctuations in the
resistance $R$. Consequently, the subgap {\it IV}-curve displays a rather well-defined peak
centered around $V = 2E_C/e$ due to the 2$e$ charge of Cooper pairs. This characteristic
feature of the {\it IV} curve provides a straightforward way to determine $E_J$ of the investigated devices of small $E_J$.

The actual downward and upward transition rates
$\Gamma_{in\downarrow}(V_C)$ and $\Gamma_{\uparrow}(V_C)$ as a
function of the collector voltage were calculated by
Zaikin and Golubev \cite{zaikin}. The Zener tunneling rate in a
resistive environment, and with the assumption $E_J \ll E_C$, is
given by
\begin{equation}
\Gamma_{\uparrow} = \frac{v}{2 \tau} \exp \left \{
-\frac{v_Z}{v-1}\left [ 1 + \frac{\left < \delta q^2/e^2 \right
>}{(v-1)^2} \right ]\right \},
   \label{uprate}
\end{equation}
and the down relaxation rate due to charge fluctuations is given
by
\begin{equation}
\Gamma_{in\downarrow} = \frac{v_Z}{\tau \sqrt{2 \pi \left < \delta
q^2/e^2 \right >}} \exp \left \{ - \frac{(v-1)^2}{2 \left < \delta
q^2/e^2 \right >}\right \},
   \label{downrate}
\end{equation}
where $v = CV_C/e$, $\tau = R_C C$ , $\left < \delta q^2 \right > = k_B C T$, and
\begin{equation}
    v_Z =  \frac{\pi^2 R_C}{8 R_Q} \left ( \frac{E_J}{E_C} \right
    )^2.
   \label{zenertunneling}
\end{equation}
The voltage $v_Z$ is related to the so called Zener break down
current by $I_Z = ev_Z/(4 \tau)$.

\subsection{BOT modeling near the onset of the bistability}

Our present model generalizes the previous analytic BOT theories \cite{hassIEEE,delahaye1} by including the effect of intra-band transitions. The circuit
schematics for the basic BOT modeling is depicted in Fig.~\ref{BOTschema} below. The basic circuit
elements are the Josephson junction, or superconducting quantum interferometric device (SQUID) geometry, at the emitter,
with a total normal state tunnel resistance of $R_{JJ}$, the
single tunnel junction at the base with the normal state
resistance $R_N$, and the collector resistance $R_C$. The BOT base
is current biased via a large resistor $R_B$ at room
temperature, but a large line capacitance $C_B$ results in an
effective voltage bias.

As required by the $P(E)$-theory, our basic modeling is valid provided $E_J P(2eV) \ll 1$.
The intrinsic relaxation is detrimental for BOT operation
and, thus, the fluctuations should be kept low by requiring that
$R_C \gg R_Q = h/4{\rm{e}}^2$. In practice, we need $R_C \gtrsim
100R_Q$ to be close to the presumed idealized operation. Experimentally, this is quite hard to realize (see Sec. III)

 Numerical analysis is needed to calculate properly the characteristics of the BOT devices near the onset of bistability. However, by introducing a phenomenological variable that describes the average number of the tunneling events $\left\langle {{N_e}} \right\rangle$ before a downward transition is triggered by the base electrons \cite{hassel04}, we may derive a rather simple description for the operation of the BOT. A value of $\left\langle {{N_e}} \right\rangle \gg 1$ is facilitated by intra-band transitions that basically maintain the bias current of the operating point. Changes in the ratio of the bias current and the triggering current can lead to significant changes in the characteristics of the BOT.

Like in the earlier analytic descriptions, the BOT emitter
current can be thought of as the result of being in
either of the two states; the Bloch oscillation state with a
time-averaged constant current and the blockaded state with zero
current,

\begin{equation}
I_E = \left \{ \begin{aligned} V_C/R_C,& \qquad \tau_{\uparrow} = 1/\Gamma_{\uparrow}\\
0,& \qquad \tau_{\downarrow} = 1/(\Gamma_{\textrm{in} \downarrow} + \Gamma_B/\left\langle {{N_e}} \right\rangle).\\
\end{aligned}\right.
   \label{collectorcurrent}
\end{equation}
The amount of time the system spends in each state is given by the
Zener tunneling rate, $\Gamma_{\uparrow}$, the intrinsic
relaxation $\Gamma_{\textrm{in}\downarrow}$, and the quasiparticle tunneling rate $\Gamma_B$; only every $\left\langle {{N_e}} \right\rangle$th of the injected base electrons is able to make a downward transition. The base current, however, flows during the
opposite times:
\begin{equation}
I_B = \left \{ \begin{aligned} 0,& \qquad \tau_{\uparrow} = 1/\Gamma_{\uparrow}\\
e \Gamma_B,& \qquad \tau_{\downarrow} = 1/(\Gamma_{\textrm{in} \downarrow} + \Gamma_B/\left\langle {{N_e}} \right\rangle).\\
\end{aligned}\right.
   \label{basecurr}
\end{equation}
From these equations we can simply derive the average emitter and
base currents
\begin{eqnarray}
\left < I_E \right > = \frac{V_C}{R_C}
\frac{\tau_{\uparrow}}{\tau_{\uparrow} + \tau_{\downarrow}}\\
   \label{avgcollectorcurrent}
 \left < I_B \right > = e
\frac{\left\langle {{N_e'}} \right\rangle}{\tau_{\uparrow} + \tau_{\downarrow}}
\label{avgbasecurrent}
\end{eqnarray}
where we have defined 
\begin{equation}
 \left\langle N_{e}^{\prime }\right\rangle =\frac{\left\langle N_{e}\right\rangle }{1+\frac{\Gamma _{in}}{\Gamma _{B}}\Ne}.
\label{neprime}
\end{equation}

By combining these two equations, we may write
\begin{equation}
\left < I_E \right > = \frac{V_C}{R_C} \frac{\tau_{\uparrow}}{e\left\langle {{N'_e}} \right\rangle} \left < I_B \right >.
   \label{avgcurrent}
\end{equation}
Now, when calculating the current gain $\beta_E = \frac{\partial \left < I_E \right >}{\partial \left < I_B \right
    >} $, $\left\langle {{N_e'}} \right\rangle$ has to be considered as a function of $\left\langle {{I_B}} \right\rangle$. Thus, we obtain
\begin{align}
    \beta_E
    = \frac{V_C}{R_C} \frac{\tau_{\uparrow}}{e\left\langle {{N_e'}} \right\rangle} -
    \frac{V_C}{R_C} \frac{\tau_{\uparrow}}{e\left\langle {{N_e'}} \right\rangle^2}
    \frac{\partial \left < N_e' \right >}{\partial \left < I_B \right>}\left\langle I_{B}\right\rangle,
    \label{betac}
\end{align}
which can equivalently be written as 
\begin{equation}
     \beta_E =  \frac{V_C}{R_C} \frac{\left < I_B \right> \tau_{\uparrow}(\tau_{\uparrow}+\tau_{\downarrow})}{e^2 \left\langle {{N'_e}} \right\rangle^2}\frac{1}{1-\beta_H}
\label{betae}
\end{equation}
with 
\begin{equation}
\beta _{H}=\frac{\tau _{\uparrow }+\tau _{\downarrow }}{\left\langle N_{e}^{\prime }\right\rangle }\frac{\partial \left\langle N_{e}^{\prime }\right\rangle }{\partial \tau _{\downarrow }}=\frac{e}{\left\langle I_{B}\right\rangle }\frac{\partial \left\langle N_{e}^{\prime }\right\rangle }{\partial \tau _{\downarrow }}.
\label{betah}
\end{equation}
When $\beta_H$ becomes equal to one, the gain diverges, which marks the threshold for bifurcation. In the regime $\beta_H >1$, two stable solutions are available and the operation becomes hysteretic as observed both experimentally and numerically. Hence, we may consider $\beta_H$ as a  parameter controlling the proximity of the bifurcation threshold.

For $\beta_H$$\rightarrow $1, we obtain a linear dependence between $\beta_E^{-1}$ and $\IB$ as given by

\begin{subequations}
\label{betaEinv}
\begin{equation} 
\beta_E^{-1}=\left[\frac{R_C}{V_C}\frac{e^2\left<N'_e\right>^2}{\tau_\downarrow(\tau_\uparrow+\tau_\downarrow) \IB^2}\big(-\IB-\epsilon \frac{\tau_\uparrow}{\tau_\downarrow} \big)\right] 
 \label{betaEinv1} 
\end{equation}
\begin{equation}
=\left[\frac{R_C}{V_C}\frac{\tau_\uparrow+\tau_\downarrow}{\tau_\downarrow}\big(-\IB +I_{B-H}\big)\right] 
 \label{betaEinv2} 
\end{equation}
\end{subequations}
where $\epsilon < 0$ is a phenomenological parameter to account for the variation of  $\partial \left<N_e'\right>/\partial \left<\tau_\downarrow\right>$ under various biasing conditions (see Appendix \ref{appendix1}).
 The latter term in the parenthesis of Eq. \ref{betaEinv1} specifies the threshold current $I_{B-H}$  for the bifurcated, hysteretic threshold. By substituting $\IB$ from Eq. \ref{avgbasecurrent} to the prefactor of Eq. \ref{betaEinv1}, $\left<N_e'\right>^2$ and $\IB^2$ term cancel each other leaving the prefactor with $(R_C/V_C)(\tau_\uparrow+\tau_\downarrow)/\tau_\downarrow$.
The detailed derivation of the analytic formulation is outlined in Appendix \ref{appendix2}.

Using a simple approximation for the variation of $\left< N_e'\right>$ with $\tau_\downarrow$, we may derive an analytic formula for the bifurcation threshold on the $E_J$ vs $\IB$ plane (see Appendix \ref{appendix1}). The
$E_J$ dependence of $I_{B-H}$ comes mainly  from Eq. \ref{uprate} which leads to the analytic form given by,
\begin{equation}
\frac{I_{B-H}}{e} \propto \frac{\Gamma_{se\uparrow}+\exp (-\kappa E_J^2)}{\sqrt{1+\Gamma_B^2/E_J^4}}
\label{crossovercurve}
\end{equation}
where the first term in the numerator, $\Gamma_{se\uparrow}$ is the upward transition rate due to  single electron tunneling, whereas the second term arises due to LZ tunneling. The parameter $\kappa$ involves all the other parameters inside the exponent of Eq. \ref{uprate}. 
This functional dependence between $I_{B-H}$ and $E_J$ in Eq. \ref{crossovercurve} is also in good agreement with the results of our numerical simulations.

The BOT behaviour described here is referred to as `normal' operation. In this configuration, the junction is initially in the upper band and a quasiparticle tunneling due to base current will bring the junction to the lowest band where it performs Bloch oscillations. This coherent oscillation will be inhibited by Zener tunneling and the system jumps back to the upper state and  the whole process is repeated again. If the sign of $V_C$ (and consequently $I_E$) is reversed the base current will induce transitions to the upper band, an operational mode that we call `inverted' operation. Since the `normal' operation is conceptually clearer we have concentrated our studies in this mode of BOT.

\section{Fabrication and measurement}
The BOT samples employed in this work were fabricated using a 20-nm-thick Ge mask on top of LOR 3B resist. Patterning of the Ge layer was performed using conventional e-beam lithography at 20 keV. After patterning, the PMMA layer was developed in MIBK:IPA (1:3) solution and subjected to a plasma etch with $CHF_4$ plasma. Finally, the LOR under the germanium was etched in oxygen plasma up to the desired extent of undercut.

Shadow angle evaporation at four different angles was employed to generate the structures consisting of three metals.
Originally, the BOT was envisioned to have a NIN junction as the base junction,
but the technique of manufacturing both SIS and NIN junctions on
the same sample is exceedingly difficult and, therefore, we opted to have a NIS
base junction instead. The SIS junction is formed of two Josephson junctions in the SQUID geometry; this facilitates tuning of the Josephson energy by magnetic flux.
The process order in the evaporation sequence was (\RNum{1}) Chromium, (\RNum{2}) Aluminium, (\RNum{3}) oxidization, (\RNum{4}) Aluminium, and (\RNum{5}) Copper. NMP or PG remover was used for lift-off. Oxidation was done in Ar:O$_2$ (6:1) mixture at 80 mTorr for 1 min.

\begin{figure}[bht]
\center \includegraphics[width = 4.4cm,height=5.4cm]{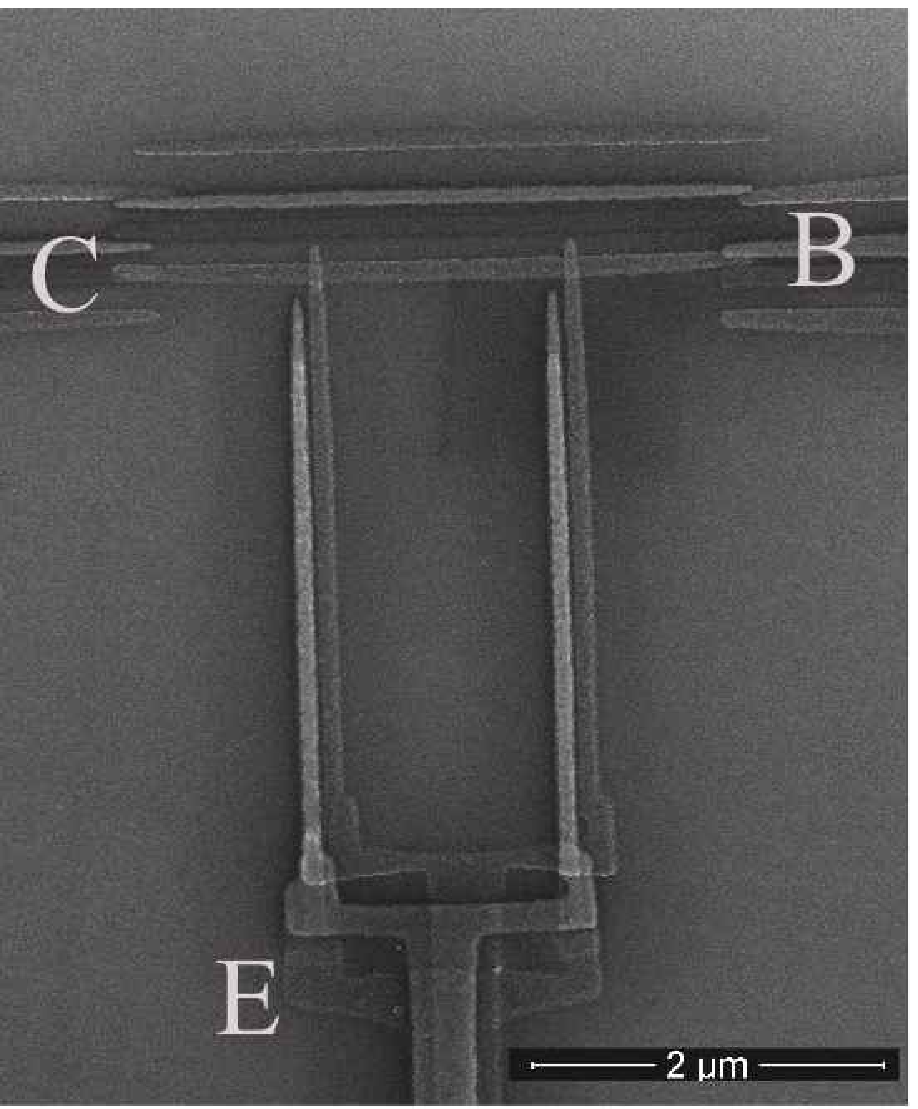}
 \includegraphics[width = 3.2cm,height=5.45cm]{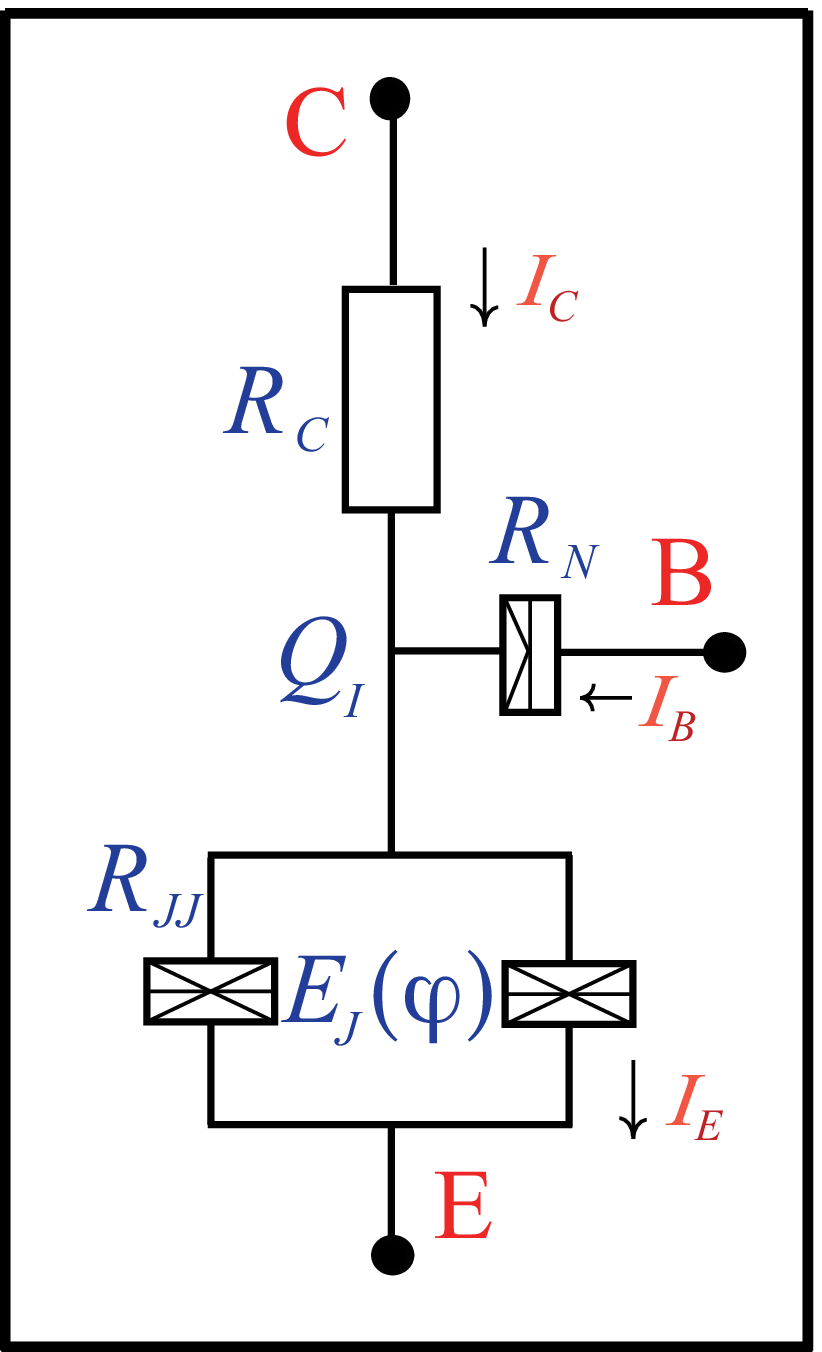}
\caption{Scanning electron micrograph of the sample (left frame) and
a schematic view of the device (right frame). In both pictures, base, emitter and
collector are marked by B, E and C, respectively. Positive directions for the currents are indicated by the arrows. The sample parameters are given in Table \ref{BOTparams}. $Q_I(t)$ is the island charge tracked in the numerical simulations.
}\label{BOTschema}
\end{figure}

A typical sample used in the present study is displayed in Fig. \ref{BOTschema}.
The area of the SIS junctions is 100 x 150 nm$^2$ each (equal areas within 10\%). The NIS junction on the base has an area $70 \times 100$ nm$^2$, roughly half of the SQUID junctions.

The measurements were done on a plastic dilution refrigerator (PDR-50) from
Nanoway Ltd. The base temperature of the refrigerator was 50 mK. The filtering in the PDR
consisted of 70 cm long Thermocoax cables on the sample
holder and 1 kOhm serial resistors at 1.5 K. In addition, micro-wave filters from Mini-circuits (BLP 1.9) were
used at top of the cryostat. 

The measurement set-up in this work was similar to that described in
Ref. \onlinecite{lindellJLTP}. The BOT base was DC current biased by a resistor $R_B$=$1-10$ G$\rm{\Omega}$, which was located at room temperature.  Voltages were measured with low noise LI-75A voltage
preamplifiers while currents were monitored using DL1211 low noise current amplifiers.

\begin{table}
\begin{tabular}{lllllllll}
\hline\noalign{\smallskip}
BOT \#&$R_{N}$&$R_{JJ}$&$R_C$& $E_J$& $E^{min}_J$&$E_C$& $\Delta$\\
\noalign{\smallskip}\hline\noalign{\smallskip}
1 & 53 & 27  & 550 & 17 & 2.7& 40 & 150 \\
2 & 75 & 21 & 305& 25 &3.3 & 60  & 165\\
\noalign{\smallskip}\hline
\end{tabular}
 \caption{BOT parameters for the measured sample. $R_{N}$ and $R_{JJ}$  are the normal
state resistances of the NIS and JJ tunnel junctions in the SQUID-loop geometry,
respectively. Resistances are given in units of k$\rm{\Omega}$ and
energies in $\mu$eV. }

\label{BOTparams}
\end{table}

The resistance values of the three circuit branches were determined at 4.2 K. Since there was a weak temperature dependence in $R_C$, we determined the actual value from $1/\sqrt{V}$ asymptote \cite{asymptote} of the {\it IV} curves measured at low $E_J$.
The maximum Josephson energy $E_J$
was calculated using the Ambegaokar-Baratoff relation which yielded $E_{J} =17$ $\mu$eV.
The flux-modified Josephson energy was obtained from the formula $  E_J(\Phi) = E_{J}\sqrt{\cos^2(\pi {\Phi / \Phi_0)}+d^2 \sin^2(\pi {\Phi / \Phi_0)}}$, where $d=\frac{E_{J_1}-E_{J_2}}{E_{J_1}+E_{J_2}}$ denotes the asymmetry in Josephson energies between the two SQUID loop junctions with $E_{J_1}$ and $E_{J_2}$, respectively. By fitting $E_J(\varphi)$ to the measured {\it IV} curves we found $d=0.15$ and 0.13 for samples \#1 and \#2, respectively.
Emitter-collector and base-emitter {\it IV} curves were employed to determine the effective energy gap $\Delta$ of the samples (see Table \ref{BOTparams}),
 which is $20-30$ $\mu$eV smaller than the bulk value $\Delta=0.18$ meV. This reduction of the gap is presumably due to the inverse proximity effect \cite{nickel} due to the chromium resistors. The spatial variation of the inverse proximity effect would also explain the larger asymmetry between the SQUID junctions than is expected due to the difference in their areas.

\section{Experimental results}

\subsection{{\it IV} characteristics}

\begin{figure}[!h]
\center \includegraphics[width = 8cm]{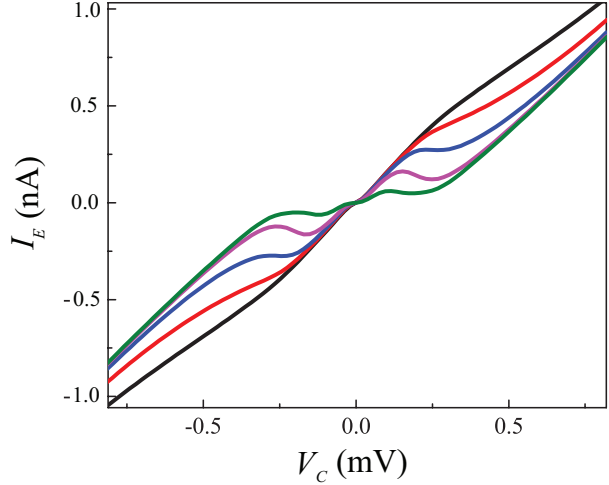}
\caption{{\it IV} characteristics of sample \#1 at a few values of Josephson coupling energy: $E_J$ = 9 $\mu$eV ({\color{black} black}), 7.3 $\mu$eV ({\color{red}red}), 6 $\mu$eV ({\color{blue}blue}), 4.5 $\mu$eV ({\color{magenta}magenta}), and 2.8 $\mu$eV ({\color{OliveGreen}green}), without base current ($I_B=0$) at $T=90$ mK.}  
\label{IVbot}
\end{figure}

{\it IV} characteristics of sample \#1 measured at a few magnetic flux values are illustrated in Fig. \ref{IVbot}: the emitter-collector current $I_E$ is recorded as a function of $V_C$ at $I_B=0$. The data clearly shows Coulomb blockade of supercurrent \cite{haviland1,kuzmin1} at all investigated values of the Josephson coupling energy. The peak in the {\it IV} in the subgap region is a signature of the inelastic Cooper pair tunneling, commonly referred to as {\it P(E)} peak (cf. Eq. \ref{pe}). The weakness of the blockade in Fig. \ref{IVbot} is assigned to the small Coulomb energy $E_C=40$ $\mu$eV, the value of which is determined from the position of the $P(E)$ peak, extrapolated to $E_J=0$. At larger bias voltage, Zener tunneling to higher bands takes place, which causes the phase fluctuation theory to break down. Our results on Zener tunneling are similar to those of Kuzmin et al. who investigated a single Josephson junction in an environment of chromium resistor \cite{kuzmin96}.

Fig.~\ref{BOTbaseCurr} demonstrates the effect of the base current on the {\it IV} of the BOT. The `normal' and `inverted' operation regions are defined by sign combinations ($V_C$,  $I_E$, $-I_B$) and ($V_C$, $I_E$, $I_B$), respectively \cite{hassel04a}. Both the `normal' and `inverted' modes of operation display a strong increase in the onset of the LZ tunneling current, which is seen as the movement of the shoulders in the $IV$ curves up to larger currents. The down-turning shoulders in the upper and lower sets of the $IV$ curves are  distinct features of LZ tunneling \cite{Schon} while the data at $I_B=0$ display only smeared bumps of these features. 
This enhancement of the LZ current suggests that effectively the energy gap between the ground and excited states is increased due to the noise induced by the current in the base junction.
The shoulders move even further apart with growing base current, which indicates an increase in the effective energy gap at the Brillouin zone boundary.
 In general, the `inverted' operation displays comparable characteristics as the normal operation, but we found
 that hysteretic behaviour appeared at smaller bias currents in the inverted regime compared with the normal operation mode; in some cases, these modes differed by a factor of four in the base current for bifurcation threshold. Nonetheless, since the normal operation is appeared to provide more clear-cut data, we concentrated our studies on this operating regime.
 
 \subsection{Gain determination}

\begin{figure}
\includegraphics[width=2.8in]{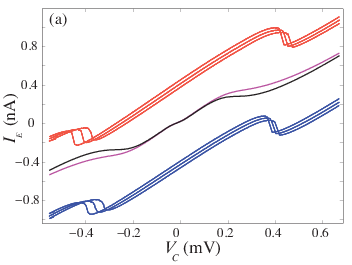}
\includegraphics[width=3.6in,height=2.4in]{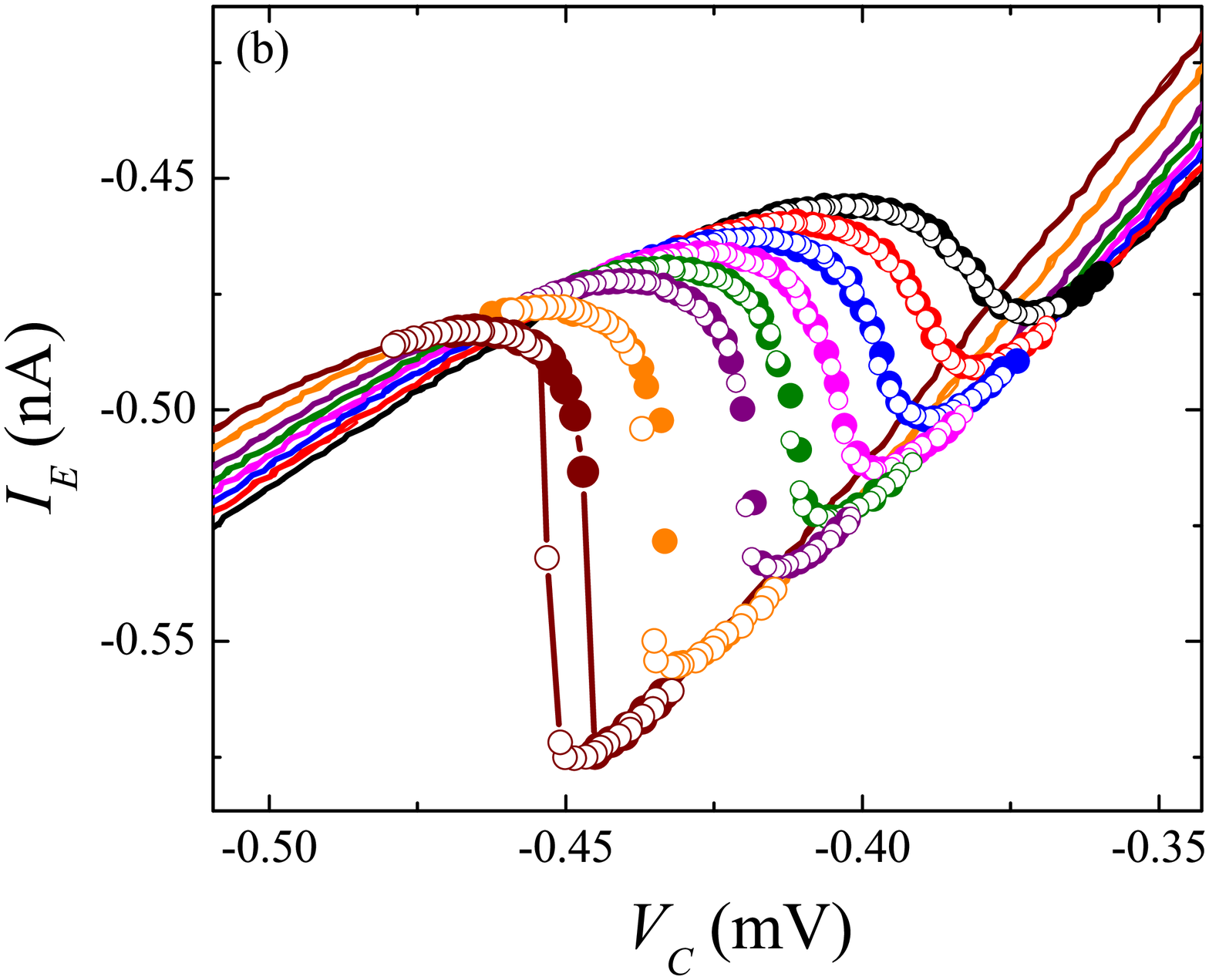} 
\caption{(a) Middle traces: {\color{magenta}magenta} and black IV curves are measured without base current at $E_J = 6.5$ and 5.8 $\mu$eV, respectively. {\color{red}Red curves}, corresponding to $E_J = 6.5$ $\mu$eV
are measured at $I_B=+0.3$, +0.34, and +0.38 nA (traces from right to left).
{\color{blue}Blue curves} have the same bias conditions as the red curves but were measured at $E_J = 5.8$ $\mu$eV.
The red  curves are offset by (+0.22 mV, +0.42 nA) for clarity, like the blue curves by (+0.22 mV, -0.42 nA).\\
(b) The normal operation region of the BOT at $E_J=7.1$ $\mu$eV with increasing $I_B$. Negative slope is the Landau-Zener tunneling regime, increases with $I_B$ and eventually the slope diverges: $I_B = +0.06$, 0.065, 0.07, 0.075, 0.08, 0.085, 0.095, and 0.105 nA (from right to left). Filled (open) circle traces are of $I_E$ when $V_C$ is swept from left (right) to right (left).The measurement temperature was at $T\sim 90$ mK.
}

\label{BOTbaseCurr}
\end{figure}

 Fig.~\ref{BOTbaseCurr}b displays a basic set of data for current gain determination in the `normal' operating region. Emitter current $I_E$ is depicted as a function of collector voltage $V_C$ at eight values of base currents $I_B$. The regime with a large negative slope marks the active bias regime of the BOT amplifier. The steepest monotonic curve (the second one from left in Fig.~\ref{BOTbaseCurr}b) has a narrow linear regime in the center of the negative slope part, the width of which amounts to about 2 pA in $I_B$. This corresponds to the maximum dynamic range in $I_B$ over which the BOT has substantial current gain at this bias point. Roughly, a change in the base current by $\Delta I_B = 2$ pA corresponds to 50 pA in $I_E$, and the current gain becomes $\beta_E=25$. Eventually, the slope of the $IV$ diverges with increasing $I_B$, after which the $IV$ characteristics become hysteretic as seen at the largest value of $I_B=0.105$ nA in Fig.~\ref{BOTbaseCurr}b. Clearly, in the BOT operation near the divergence point, the dynamic regime is inversely proportional to the current gain.

 We have checked that hysteresis does not depend on the value of the current bias resistor in the range $10^8-10^{10}$ $\Omega$. Moreover, we have performed simultaneous transconductance $g_m=\frac{\Delta I_E}{\Delta V_B}$ and current gain measurements to determine the input impedance of the BOT $Z_{in}=\frac{\Delta V_B}{\Delta I_B}=\frac{\Delta V_B}{\Delta I_E} \times \frac{\Delta I_E}{\Delta I_B}=\frac{\beta_E}{g_m}$. We find that the input impedance diverges at the same point as the gain.

 According to basic BOT theories \cite{hassel04,delahaye2,delahaye1}, the current gain is independent on the base current.
 However, the situation changes near the bifurcation point. This is because there can be two different kinds of base current components: one comprising of tunneling events causing interband transitions (the only component in the traditional BOT base current) and another one leading only to intraband events. Only the interband transitions lead to gain in the BOT while the intraband transitions are to maintain the bias current. As the base current grows, the ratio of these two current components may change with increasing $I_B$ which leads to current dependence of the gain and, eventually, to the diverging behavior when approaching the bistability point. Hence, the observed strong increase in $\beta_E$ with increasing base current is a sign of the operation near the bifurcation point where the gain grows according to Eq. \ref{betaEinv}.

\begin{figure}
\includegraphics[width = 8cm,height=6cm]{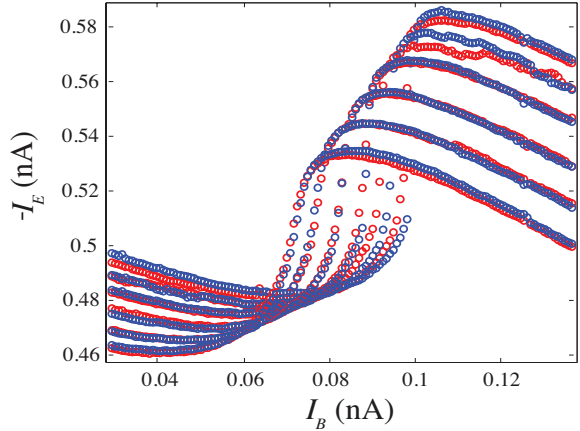}
\caption{Measurement of the current gain by tracing $I_E$ vs $I_B$ at $E_J$= 7.1 $\mu$eV. The steepest slope yields the operating point with the largest current gain $\beta_E$ at the corresponding collector voltage $V_C$. Traces were measured at $T = 90$ mK using $V_C=-0.443$, -0.429, -0.419, -0.410, and -0.401 mV (traces from right to left). Different signs of $I_E$ and $I_B$ correspond to the regime of `normal' operation. Red (purple) traces are for growing (decreasing) sweep of $I_B$.}\label{IcIb}
\end{figure}

 In addition to the analysis of data like in Fig.~\ref{BOTbaseCurr}b, we have measured the current gain using traces of $I_E$ vs $I_B$ as illustrated in Fig. \ref{IcIb}. The figure displays data  at five different values of the $V_C$ at $E_J=7.1$ $\mu$eV. The current gain is calculated from the negative slope of $I_E(I_B)$: $\beta_E = - \frac{\Delta I_{E}}{\Delta {I_B}}$. The maximum of the decaying slope yields the optimum current gain, which we determined as an average of the up and down $I_B$ sweeps. At large gains, there was often a rather large difference ($\sim$ factor of 2) between the gains of up and down sweeps. In such cases, we considered the operation of the BOT bistable at this bias point and disregarded the larger gain value. We performed $I_E(I_B)$ measurements at different values of $E_J$  and picked out a single gain value $\beta_E(I_B)=\left.- \frac{\Delta I_{E}}{\Delta {I_B}}\right|_{max}$ from each of the data sets. When $\beta_E > 50$, the gain determinations became problematic because of $1/f$ noise and creep in the measurement, which gradually took the device out of the linear regime during the gain determination.

Fig. \ref{gain} depicts data on the inverse of $\beta_E$ vs $I_B$ which were obtained from the analysis of $I_E(I_B)$ scans performed in the range with $E_J= 4.6- 10.5$ $\mu$eV. Plotting $\beta_E^{-1}$ makes the analysis of the diverging gain regime simpler and allows us to examine the vicinity of the bifurcation point where we are supposed to have $\beta_E^{-1}(I_B) \rightarrow 0$. Experimentally, the problem in this region arises because the dynamic range becomes zero and measurements without noise induced smearing become impossible. Nevertheless, the data in Fig. \ref{gain} display how the critical regime is approached at $\beta_E^{-1}>0.02$ which corresponds to our highest reliable gain values. All of the data at small values of $\beta_E^{-1}$ are seen to show a nearly linear dependence on $I_B$, especially at large values of Josephson energies. At our smallest value of $E_J= 2.7$ $\mu$eV (sample \#1), we could not reach the bistable regime at all. The theoretical dependence for $\beta_E^{-1}(I_B)$, illustrated by red curves in Fig. \ref{gain}, were obtained by fitting Eq. \ref{betaEinv1} to the data just near the divergence point, as required by its regime of validity. 

\subsection{Bifurcation threshold}

\begin{figure}
\includegraphics[width=83mm,height=60mm]{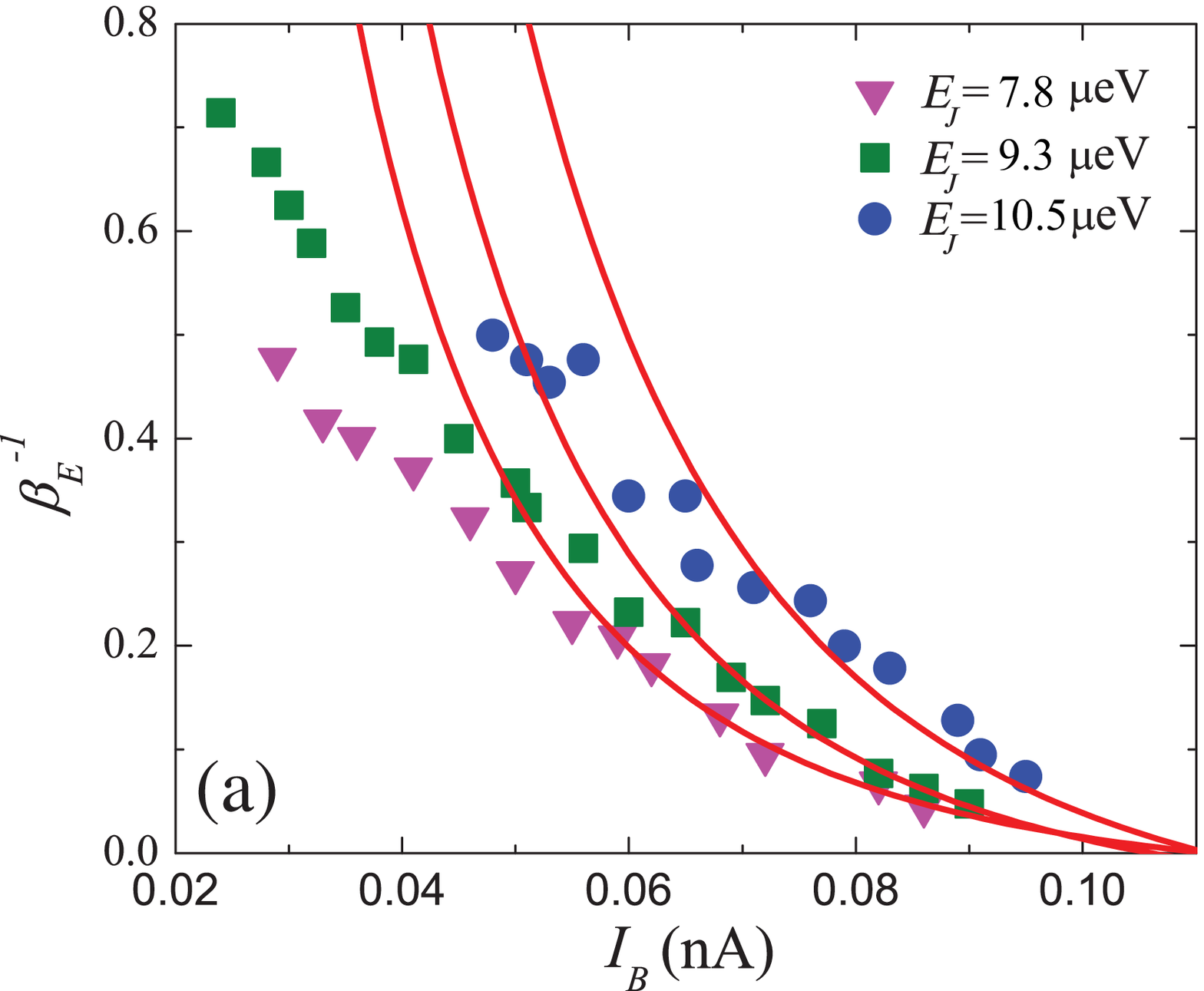}
\includegraphics[width=83mm,height=59mm]{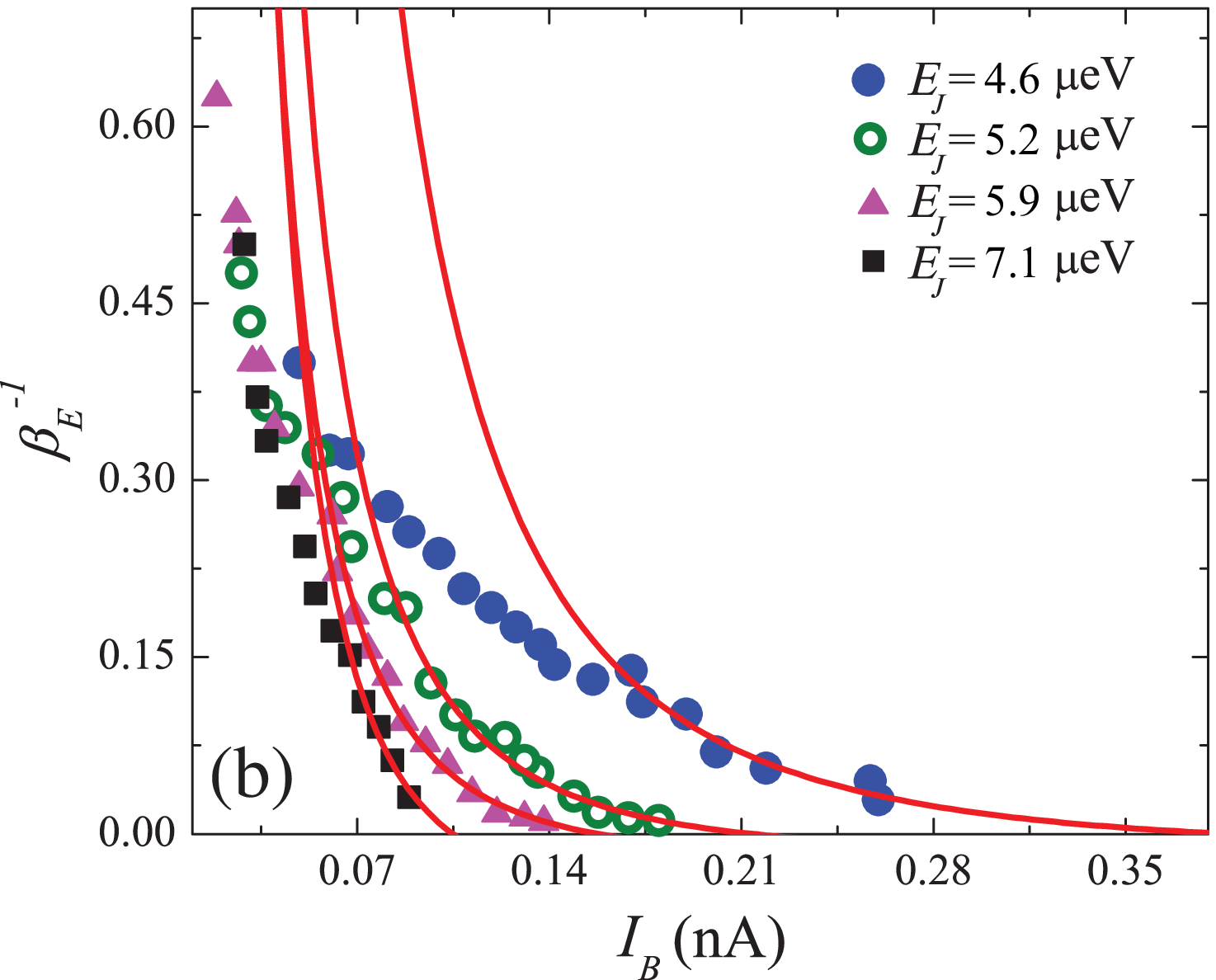}
\caption{Inverse gain $1/\beta_E$ as a function of bias current $I_B$. Each data point was obtained from a $I_E$ vs $I_B$ sweep illustrated in Fig. \ref{IcIb}. The solid curves were obtained using Eq. \ref{betaEinv} fitted to the highest $I_B$ quartile fraction of the data sets (1st-8th lowest $\beta_E^{-1}$ values).
} \label{gain}
\end{figure}

\begin{figure}
\includegraphics[width = 8cm]{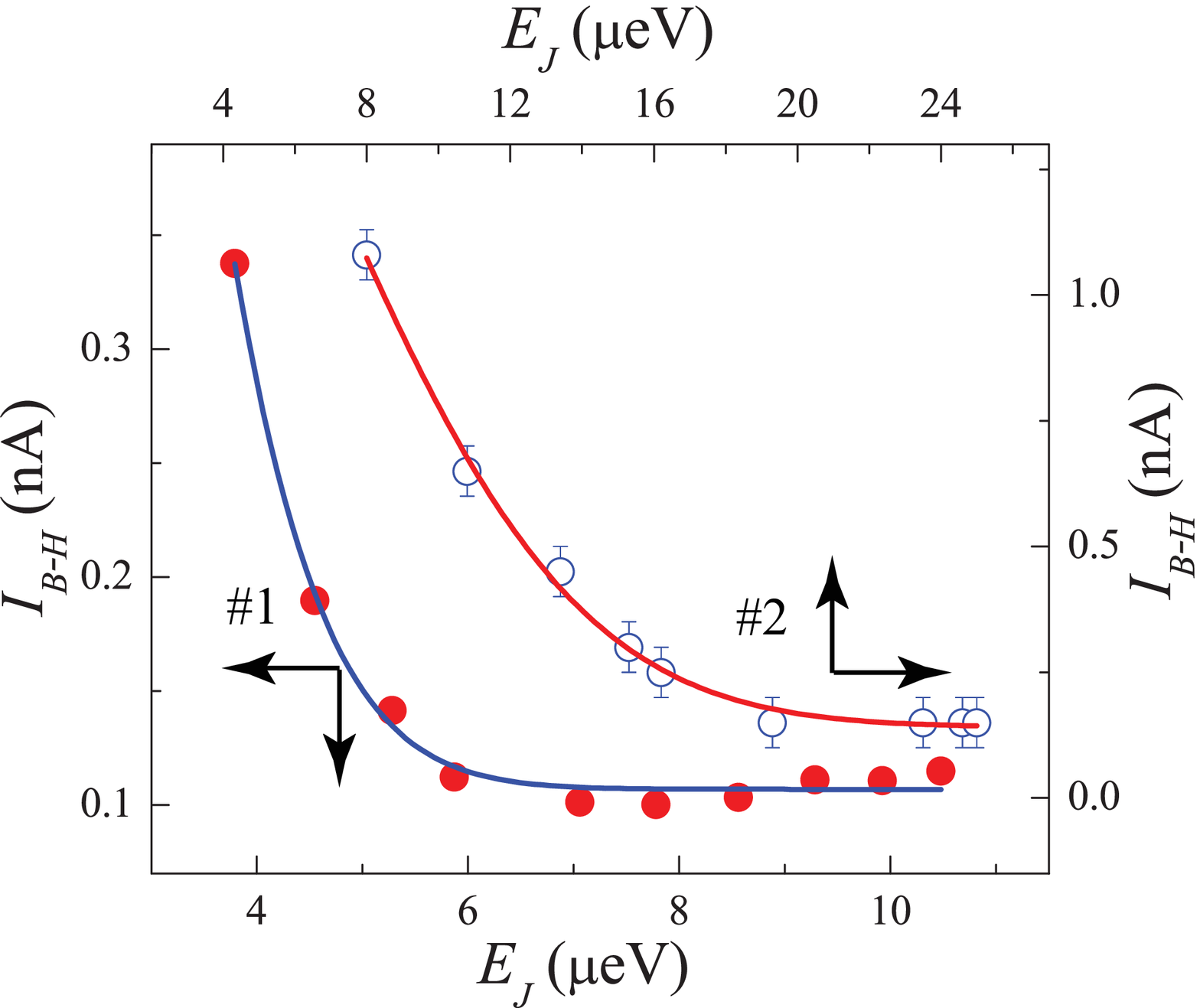}
\caption{Bifurcation threshold on the $E_J$ vs $I_B$ plane. Red(filled)  and blue(open) circles denote the $I_{B-H}$ values for the sample \#1 and \#2, respectively. Solid curves display the theoretical dependence from Eq. \ref{crossovercurve}.
} \label{Phase}
\end{figure}

The experimentally determined values of $I_{B-H}$ for the bifurcation point are plotted in Fig. \ref{Phase} on the $I_{B-H}-E_J$ plane. The plot was generated from the fits in Fig. \ref{gain} by selecting the points of $\beta_E^{-1}(I_{B-H})=0$. Fig. \ref{Phase} indicates that the onset of bistability is nearly independent of base current at large values of $E_J$, while a steep increase in $I_{B-H}$ is observed below $E_J = 6$ $\mu$eV. The observed behavior is quite well reproduced by our phenomenological formula Eq. \ref{crossovercurve} which is depicted by the solid curve.

For sample \#2, we found a similar threshold curve which indicates that the bifurcation behaviour and its dependence on $E_J$ is a fundamental characteristic of Bloch oscillating transistor. The bifurcation threshold curves for both the samples are depicted in Fig. \ref{Phase}. For sample \#2, the bifurcation threshold current is higher than that for sample \#1. From the fitted curves we found that $\kappa$ (see Eq. \ref{crossovercurve}) for sample \#1 is higher than for sample \#2 which comes from the fact
that $\kappa$ contains term $R_C$  which is higher in sample \#1 than in sample \#2. The smaller base current needed for bifurcation for sample \#1 than sample \#2 is accounted from the fact that $\beta_H$ for sample \#1 is higher than for sample \#2.
Absence of bifurcation was observed in both sample \#1 and \#2 at their respective lowest $E_J$ values.

\begin{figure}[ht]
 \subfigure{
\includegraphics[width=6cm]{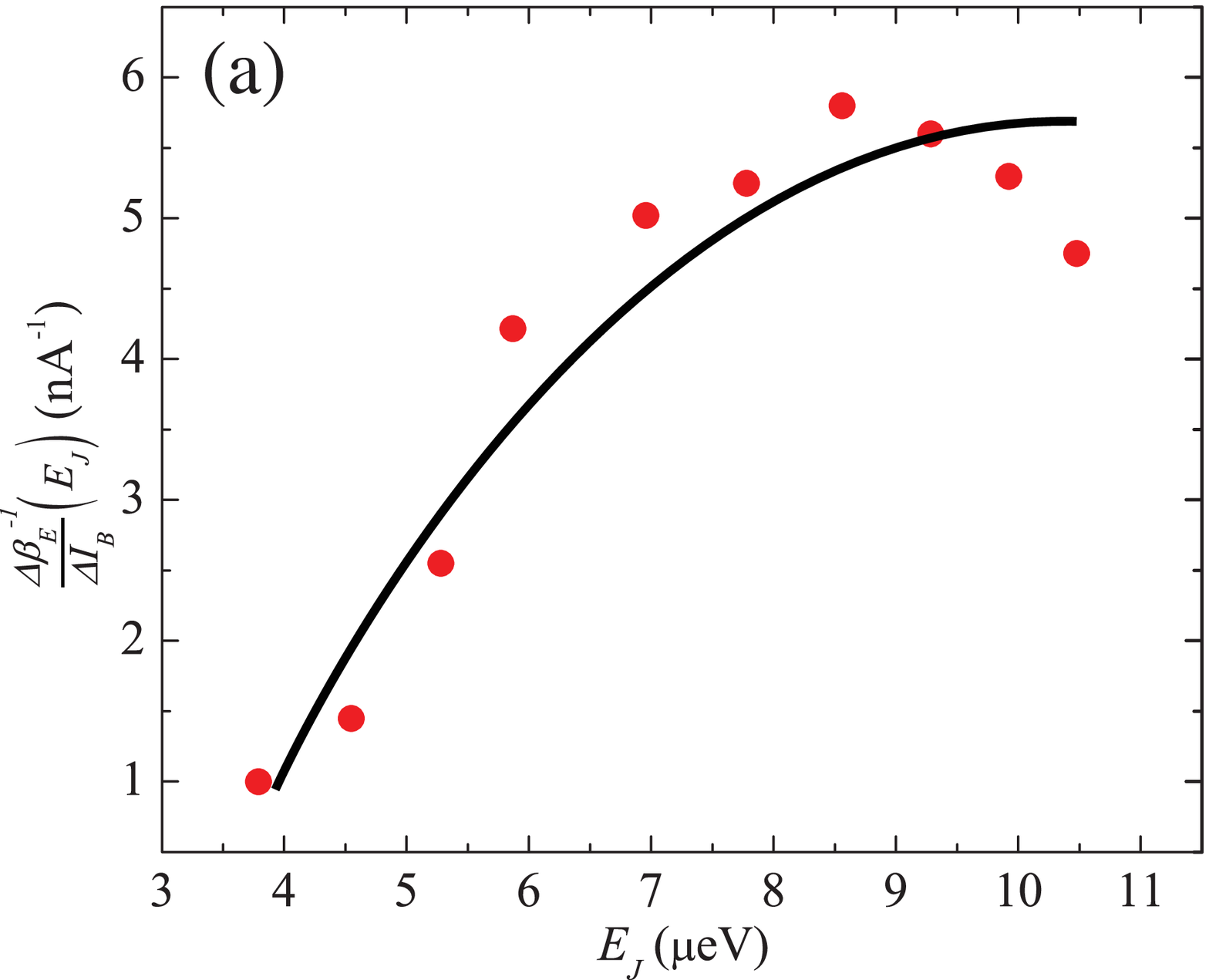}
\label{figure7a}
}
 \subfigure{
\includegraphics[width=6cm]{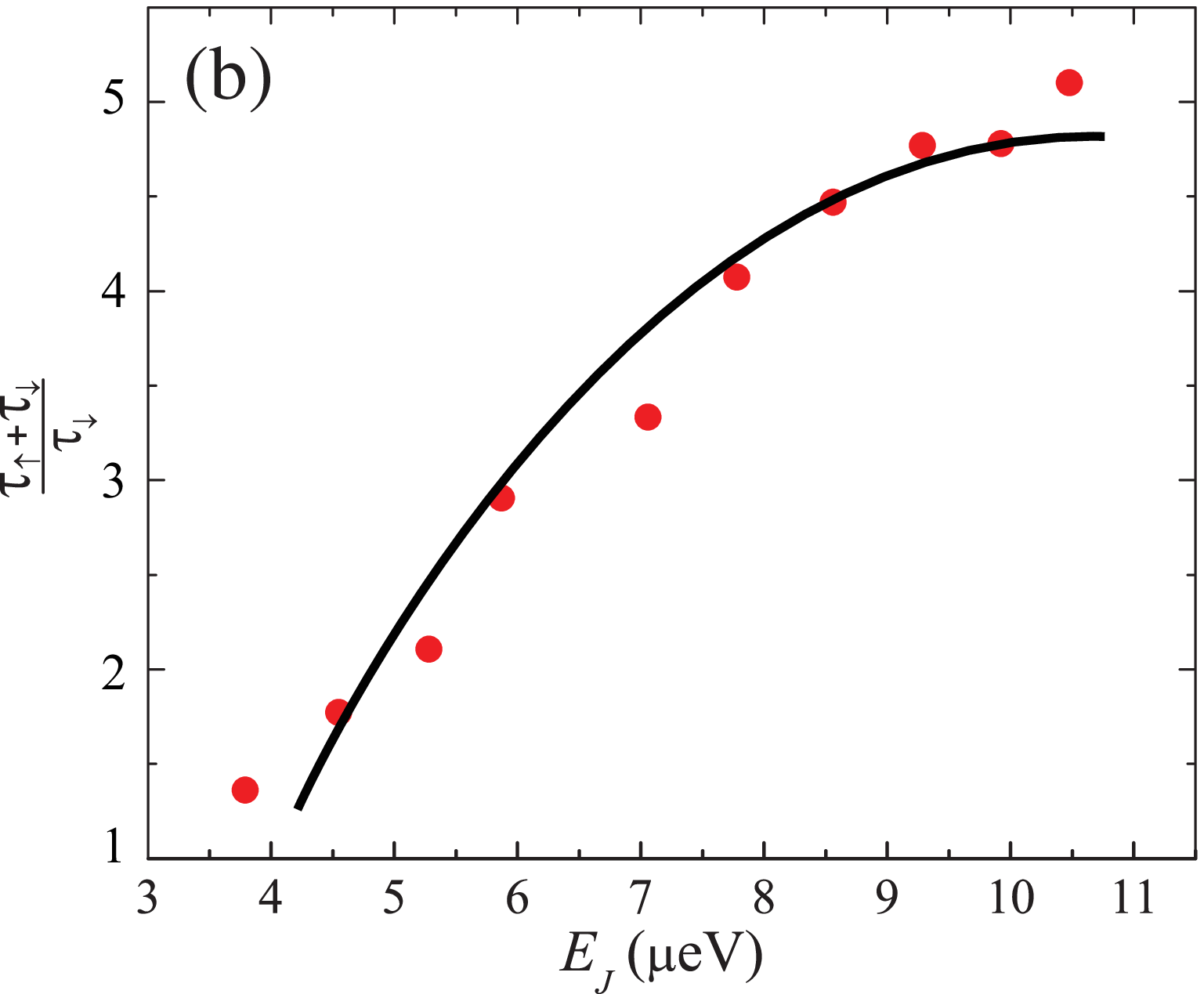}
\label{figure7b}
}
\label{slopeej} 
\caption{(a) $\frac{\Delta \beta_E^{-1}}{\Delta I_B}$ vs $E_J$  plotted near the bifurcation threshold. Each point at different $E_J$ was derived from fits in Fig. \ref{gain}. (b) $\frac{\tau_\uparrow+\tau_\downarrow}{\tau_\downarrow}$, deduced from experimentally determined $\frac{\IE}{V_C/R_C}$, is plotted vs $E_J$. Solid curves are seen to display the similar character but the theoretical estimate falls short by 70\% from the change in Fig. \ref{figure7a}}     
\end{figure}

The rate ($\frac{\Delta\beta_E^{-1}}{\Delta I_B}$) at which 1/$\beta_E$ reaches the bifurcation point depends on $E_J$ as seen from Fig. \ref{figure7a}. Initially, the slope increases rapidly with $E_J$ upto 6 $\mu$eV, while between 6 - 10.5 $\mu $eV the slope appears to be saturated. In this region, the bifurcation threshold current is almost indepedent of $E_J$ (cf. Fig.  \ref{Phase}). The slope variation over the whole range of $E_J$ amounts to a factor of 5.5. Hence, the variation of  $E_J$ does not necessarily change $\beta_E$ strongly, which is a desirable property concerning $1/f$ noise due to critical current fluctuations.

Theoretically, the rate ($\frac{\Delta\beta_E^{-1}}{\Delta I_B}$)  is hard to evaluate from Eq. \ref{betaEinv}. The significant prefactor of Eq. \ref{betaEinv2} contains two terms : $V_C/R_C$  and $(\tau_{\downarrow}+\tau_\uparrow)/\tau_{\downarrow}$. 
We have analyzed how $\frac{\Delta\beta_E^{-1}}{\Delta I_B}$ varies theoretically with $E_J$ for sample \#1 by estimating the factor $\frac{R_C}{V_C}\frac{\tau_\uparrow+\tau_\downarrow}{\tau_\downarrow}$. By using Eq. \ref{avgcollectorcurrent}, we can relate $\frac{\tau_{\downarrow}+\tau_{\uparrow}}{\tau_{\downarrow}}$ to the experimentally determined quantity $\frac{\IE}{V_C/R_C}$, while $V_C$ is obtained from the bias voltage, which is increased by 60\% over the range of $E_J=3.9-10.5$ $\mu$eV.
 In Fig. \ref{figure7b}, we show the variation of $\frac{\tau_{\downarrow}+\tau_{\uparrow}}{\tau_{\downarrow}}$ with $E_J$ as determined for the maximum current $\IE$ at the sub-gap $I(V)$ peak.
Together, these opposing contributions result in a change by a factor of 3.3 in $\frac{\Delta\beta_E^{-1}}{\Delta I_B}$, which falls slightly short from the observed factor of 5.5 in Fig. \ref{figure7a}. Hence, we can conclude that our simple model explains the rate of approach towards the bifurcation threshold with fair extent.

We have also tried to determine the ratio of inter- and intra-band transitions which is governed by $\Ne$.
According to Eqs. \ref{avgcollectorcurrent} and \ref{neprime} there is the following relation between the base current and $\Ne$ : $\IB=\frac{e \Ne^2}{\Gamma_B}\frac{\tau_\downarrow}{\tau_\uparrow+\tau_\downarrow}$. With the approximation, $\IB \propto \Gamma_B$,  we can conclude that $\Ne \propto I_{B-H} \sqrt\frac{\tau_\uparrow+\tau_\downarrow}{ \tau_\downarrow}$. From Fig. \ref{figure7b} the increase in $\sqrt\frac{\tau_\uparrow+\tau_\downarrow}{ \tau_\downarrow}$ is $\sim$ 2 whereas from Fig. \ref{Phase}  the decrease in $I_{B-H}$ for sample \#1 is $\sim$ 4. Thus, we can conclude that $\Ne$ goes down with increasing $E_J$. But unfortunately, we cannot determine the exact number of $\Ne$ from this analytical formulation.

In our numerical analysis, we have considered the circuit model used by Hassel et. al \cite{hassel04a},  and modified it for current bias configuration. Here we have calculated the island charge as a function of time by taking into account three contributions: charge relaxation through $R_C$ together with the tunnel current through the emitter and base junctions, respectively. The tunnel currents through these junctions are calculated using time dependent $P(E)$ - theory \cite{hassel04a}. In the simulation, $P(E)$ is calculated numerically by considering only the real part of the environmental impedence. The simulation-run-time was chosen longer than the time constant due to $R_B$ and the capacitance from base to ground so that the steady state was reached properly.
Moreover, we monitored the tunneling events on the island with time, which clearly revealed the Bloch oscillating state and its transition to the higher band. By counting the number of tunneling events when the system undergoes a change from higher band to lower band we could calculate $\Ne$ from the simulation.

\section{Discussion and conclusions}

According to the basic theory of the BOT operation, the gain should depend exponentially on $E_J/E_C$ via the tunneling rates $\Gamma_{\uparrow}$ and $\Gamma_{in\downarrow}$ \cite{delahaye1} when $E_J \ll E_C$. In this small-$E_J$ limit, the energy gap between the first two bands is small, which facilitates the use of the up- and down transition rates (Eqs. \ref{uprate} and \ref{downrate}) from perturbation theory.
In our experiments we are well in this limit, which has not been the case in many of the previous measurements, for example, in Ref. \onlinecite{delahaye2} the maximum gain of $\beta_E=35$ was achieved for $E_J/E_C = 3.4$. A current gain of $\beta_E$ = 25 was reported for low $E_J/E_C = 0.3$ in Ref. \onlinecite{reneapl}. 
In our present paper, we have observed a large current gain of $\sim$ 50 even at $E_J/E_C = 0.1$.
The estimates from the basic theory \cite{delahaye1} amount to $\beta_E = 4.8 - 7$ for $E_J= 5-11.8$ $\mu$eV, well below the measured values. Moreover, we did not observe any variation of the maximum gain with $E_J$, which, together with magnitude of $\beta_E$, is consistent with the operation near the bifurcation point where the main gain mechanism has a different origin than in the regular BOT operation.

In the operating regime near the bifurcation point, the base current is a combination of a working point current, not inducing inter-band transitions, and a significantly smaller part that leads to transitions, the ratio of these two currents being given by the parameter $\langle N_e \rangle$. In our phenomenological modeling with large number of intra-band transitions, the current gain is simply related to $\left\langle N'_e \right\rangle$ and the upward tunneling rate $\Gamma_{\uparrow}$ (see Eqs. \ref{neprime} and \ref{avgcurrent}). Hence, a large current gain is expected when approaching a regime where there are two stable solutions for the base current with different values for $\left\langle N_e \right\rangle$. When $\langle N_e \rangle$ is large, then almost all of the current is used to just keep the operating point. In our simulation we find a factor of 15 change in $\left<N_e\right>$ over the measured range but
unfortunately, our analysis is not able to yield absolute numbers for ${\left\langle {{N_e}} \right\rangle}$ from our measured data.

When comparing our findings with the numerical work of Hassel and Sepp\"a \cite{hassel04,hassIEEE}, we find a weaker overall dependence of the device performance on the sample parameters and biasing parameters than was found in the simulations. The weaker overall dependence may, of course, be valid only for the regime of the sample parameters/device configurations that were investigated in the present work. Nevertheless, the weaker parameter dependence is an important factor that contributes to the success of the simple phenomenological modeling that we have employed. The weaker overall changes may also indicate that there is external noise present in the measurements and our results should be compared with simulations performed at a higher effective temperature. Furthermore, as an example of the differences, let us point out that if we take the hysteresis parameter from Ref. \onlinecite{hassel04}, $\beta_H=0.02\left (\frac{R_C}{R_N}\right )^2 \exp\left ( \frac{\pi e^2 R_C}{16\hbar}\left ( \frac{E_J}{E_C}\right)^2\right)$, we find that our sample \#1 should be bifurcated at all base currents ($\beta_H \sim 2.4-11.5$).  
We think that the absence of bifurcation at $I_B=0$ with $\beta_H$ well above 1 indicates the necessity to add  a capacitance in parallel to $R_C$ into the simulations, which would take into account the parasitic capacitance component on the sample chip. This parasitic capacitance will influence the Coulomb blockade at large frequencies, at which it will reduce the real part of the impedance seen by the Josephson junction.

In conclusion, we have investigated the dynamics and modeling of the BOT when approaching a bifurcation point governed by intricate interband transition dynamics. Our results present the first experimental analysis on the behavior of the BOT in the regime where its behavior is fully governed by switching dynamics with the rates imposed by the biasing conditions. We have reached record-large current gains even though the device was operated just at small Josephson coupling energies $E_J=2.7 - 10.5$ $\mu$eV. We have mapped a cross-over transition diagram on the $E_J$ vs $I_B$ plane and compared its shape to analytic modeling, where the intraband transitions are included in terms of a phenomenological parameter $\Ne$. The same modeling was also successfully applied to describe how the current gain diverges as a function of the base current $I_B$. 
Our findings are consistent with the gain divergence as 1/(1-$\beta_H$) where the hysteresis parameter $\beta_H$ is only weakly dependent of $E_J$. The weak dependence makes this regime attractive for application where large current gain is needed at low frequencies.

\section*{Acknowledgements}
We acknowledge fruitful discussions with Antti Manninen, Mikko Paalanen and Heikki Sepp\"a. We acknowledge Micronova cleanroom facilities for fabrication of our studied samples.
Financial support by Academy of Finland, Technology Industries of Finland Centennial Foundation and TEKES is gratefully acknowledged.

\appendix
\section{Analytical derivation of threshold curve}
\label{appendix1}
It is difficult to obtain an analytic expression for the derivative $\partial \left<N'_e\right>/\partial \tau_\downarrow$ and hence, we had to be satisfied with crude approximations.
Eq. \ref{betah} specifies the relation between $\beta_H$ and the derivative $\partial \left<N'_e\right>/\partial \tau_\downarrow$ as follows: $\beta_H=\frac{e}{\left< I_B \right >}\frac{\partial \left<N'_e\right>}{\partial \tau_{\downarrow}} $.
The partial derivative of $\left<N'_e\right>$ with respect to $\tau_{\downarrow}$ can be approximated as
\begin{equation}
\frac{\partial \left<N'_e\right>}{\partial \tau_{\downarrow} }=\frac{\left<N'_e\right>}{ \tau_{\downarrow} }+\epsilon,
\label{correctionterm}
\end{equation}
where $\epsilon<0$ is a phenomenological correction term. 
In order to determine the variation of $\left < N'_e \right >$  with $\tau_{\downarrow}$ explicitly,  we follow an interpolative approach using 
\begin{equation}
\left < N'_e \right >=\left[1+(\Gamma_B\tau_{\downarrow})^2\right]^{1/2},
\label{neinter}
\end{equation}
which agrees with the limits; when $\tau_\downarrow$ is short $\left<N_e'\right>$ approaches 1 and $\left<N_e'\right> \simeq \Gamma_B \tau_\downarrow$ when $\tau_\downarrow$ is long.
Hence,
\begin{equation}
\frac{\partial \left<N'_e\right>}{\partial \tau_{\downarrow}} =\frac{\sqrt{1+(\Gamma_B\tau_{\downarrow})^2}}{\tau_{\downarrow}},
\end{equation}
which leads to,
\begin{eqnarray}
\epsilon&=&\frac{\partial \left<N'_e\right>}{\partial \tau_{\downarrow} }-\frac{\left<N'_e\right>}{ \tau_{\downarrow} } \nonumber \\
&=&-\frac{1}{\tau_\downarrow \sqrt{1+(\Gamma_B\tau_\downarrow)^2}}.
\label{epsl}
\end{eqnarray}


Using Eq. \ref{correctionterm},  we can write for the hyteresis parameter,
\begin{eqnarray}
\beta_H=\frac{e}{\left< I_B\right>}\frac{\partial\left<N'_e\right>}{\partial \tau_{\downarrow}}
\\
=\frac{e}{\left<I_B\right>}\left[ \frac{\left<N'_e\right>}{\tau_\downarrow} +\epsilon \right]
\label{betahapprox}
\end{eqnarray}
Using the expression of $\IB$ from Eq. \ref{avgcollectorcurrent};
\begin{equation}
\IB=\frac{e \left< N'_e\right>} {\tau_{\downarrow}\left (1+\frac{\tau_{\uparrow}}{\tau_{\downarrow}}\right)}, \nonumber
\label{ib_correctionterm}
\end{equation}
and substituting it with $\frac{ \left< N'_e\right>} {\tau_{\downarrow}}$ into Eq. \ref{betahapprox}, we get the following equation

\begin{eqnarray}
\frac{ I_{B-H}}{e}&\simeq&-\epsilon \frac{\tau_\downarrow}{\tau_\uparrow}
\end{eqnarray}
at $\beta_H=1$.
By inserting $\epsilon$ from Eq. \ref{epsl}, we obtain an analytic expression for the bifurcation curve,
\begin{equation}
I_{B-H}= \frac{1}{\tau_\uparrow}\frac{1}{\sqrt{1+(\Gamma_B\tau_\downarrow)^2}}
\end{equation}
The upward transition rate $1/\tau_\uparrow$, depends on both LZ tunneling $\Gamma_{\uparrow}$ and single electron tunneling ($\Gamma_{se\uparrow}$). Single electron events induced by the base current were found to be important in the simulated time traces of the island charge at high $E_J$ values. Hence,
\begin{equation}
\frac{1}{\tau_\uparrow}=\Gamma_{\uparrow}+\Gamma_{se\uparrow}
\end{equation}
According to LZ tunneling ($\Gamma_{\uparrow}$) (cf. Eq. \ref{uprate})
\begin{equation}
\Gamma_{\uparrow} \propto \exp(-\kappa E_J^2),
\end{equation}
where the parameters inside the exponent are absorbed in $\kappa$. Thus, we arrive at an exponential dependence of $I_{B-H}$ with $E_J^2$: 
\begin{equation}
\frac{I_{B-H}}{e} \propto \frac{\Gamma_{se\uparrow}+\exp (-\kappa E_J^2)}{\sqrt{1+\Gamma_B^2/E_J^4}}, \nonumber
\label{ibhfit}
\end{equation}
which is Eq. \ref{crossovercurve} in the main text. The effect of single electron tunneling reflects on the bifurcation threshold curve through the saturation of $I_{B-H} (\neq 0)$ at higher $E_J$ values.

\begin{figure}[h!]
\includegraphics[width = 8cm]{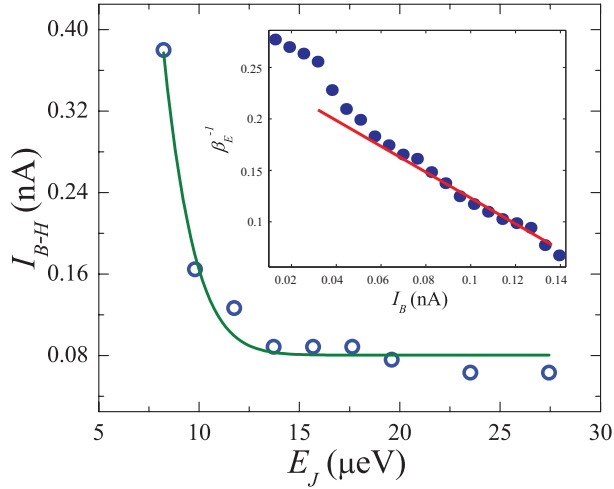}
\caption{Bifurcation threshold on the $E_J$ vs $I_B$ plane obtained from the simulation. The solid curve is the analytic dependence from Eq. \ref{crossovercurve}. The inset shows the dependence of $\beta_E^{-1}$  on $I_B$ obtained from the numerical simulation at  $E_J$=10 $\mu$eV; the fitted line indicates $I_{B-H}$ = 0.16 nA.} \label{Phase_simu}. 
\end{figure}

 We have used Eq. \ref{crossovercurve} to fit the bifurcation threshold diagram. We find good agreement with both experimental and simulated data (cf. Figs. \ref {Phase} and  \ref{Phase_simu}). Though our simulated threshold values deviate from our experimental $E_J$s, we find similar functional tendency in the curves. Both in the simulation and experiment we observe that the bifurcation takes place  earlier in `inverted' regime than in the `normal' operation. In the inset of Fig. \ref{Phase_simu}, we display a calculated $\beta_E^{-1}$ vs $I_B $ plot at $E_J$=10 $\mu$eV. In the simulation,  we also found a minimum $E_J$ below which there is no hysteresis. Hence, we can conclude that our simulation quite well explains the experimental findings. 

\section{$I_B$ vs $\beta_E$}
\label{appendix2}
According to Eq. \ref{betae}
\begin{equation}
\beta_E^{-1}=\frac{R_C}{V_C}\frac{e^2 \left<N'_e\right>^2}{\tau_\uparrow(\tau_\uparrow+\tau_\downarrow)}\frac{1}{\IB}\left(1-\beta_H\right) 
\end{equation}

By inserting $\beta_H$ from Eq. \ref{betahapprox}, we reach the approximate form of $\beta_E$ near the hysteresis point:
\begin{eqnarray}
\beta^{-1}_E&=&\left[\frac{V_C}{R_C}\frac{\IB}{e^2}\frac{\tau_{\uparrow}(\tau_\uparrow+\tau_\downarrow)}{\left<N'_e\right>^2}\frac{1}{1-\frac{e}{\IB}(\frac{e\left<N'_e\right>}{\tau_{\downarrow}}+\epsilon)}\right]^{-1} \nonumber \\ \nonumber
&=&\left[\frac{R_C}{V_C}\frac{\tau_\uparrow+\tau_\downarrow}{\tau_\downarrow}\big(-\IB +I_{B-H}\big)\right], 
\label{fullform_gaininv}
\end{eqnarray}
where $I_{B-H}$ denotes the bifurcation threshold current.
 The above formulation is valid only in the vicinity of the divergence point, where the dominant change in $\beta_E^{-1}$ can be viewed as linear in $\IB$.

\end{document}